\documentclass[prd,twocolumn,unsortedaddress,superscriptaddress]{revtex4-2}
\usepackage{graphicx}
\usepackage{epsfig,color}
\usepackage{amsmath,amssymb}
\usepackage{bm}
\usepackage{hyperref}

\begin{document}
	
	\title{Approximating Relativistic Quantum Field Theories with Continuous Tensor Networks}
	
	\date{\today}
	
	\author{Tom Shachar}
	\affiliation{Racah Institute of Physics, The Hebrew University of Jerusalem, Jerusalem 91904, Givat Ram, Israel.}
	
		\author{Erez Zohar}
	\affiliation{Racah Institute of Physics, The Hebrew University of Jerusalem, Jerusalem 91904, Givat Ram, Israel.}
	
	\begin{abstract}
		We present a continuous tensor-network construction for the states of quantum fields called cPEPS (continuous projected entangled pair state), which enjoys the same spatial and global symmetries of ground-states of relativistic field theories. We explicitly show how such a state can approximate and eventually converge to the free field theory vacuum and suggest a regularization-independent way of estimating the convergence via a universal term in the fidelity per-site. We also present a detailed bottom-up construction of the cPEPS as the continuum limit of the conventional lattice Projected Entangled Pair State (PEPS).
	\end{abstract}
	
	\maketitle
	
	\section{Introduction}
		
		Tensor network states have proven to be a prolific theoretical and numerical framework for  advancement in the understanding of many-body quantum systems. These states, which are constructed by contractions of local tensors, capture very well the physics of ground-states, low lying excited states and thermal states of local Hamiltonians \cite{cirac_renormalization_2009,verstraete_matrix_2008,schollwock_density-matrix_2011,orus_practical_2014,cirac2021matrix}. By construction, they have the relevant physical properties, including the right entanglement structure (area law \cite{RevModPhys.82.277}), possibility to encode symmetries - both global \cite{sanz_matrix_2009,molnar_normal_2018} and local \cite{haegeman_gauging_2015,zohar_building_2016,kull_classification_2017}, topological properties \cite{schuch_peps_2010}, and, of course, a large set of numerical methods applicable to different scenarios and purposes - e.g. \cite{white_density_1992,vidal_efficient_2003,verstraete_matrix_2004,zwolak_mixed-state_2004,jordan_classical_2008} and more. These approaches, including the $1+1$ dimensional Matrix Product States (MPSs) and their higher dimensional extension PEPS (Projected Entangled Pair States) \cite{cirac2021matrix}, are almost entirely restricted to lattice models, with very few extensions, so far, to quantum field theories defined in the continuum.
		
	Originally, tensor network state methods were mostly applied to condensed matter problems. In the recent years, however, there has been a growing effort in applying them also to high energy physics, in particular for the study of lattice gauge theories. That includes the benchmarking of known results, as well as computations go beyond the state of the art of conventional, Monte-Carlo methods (see, e.g., \cite{banuls_review_2020} for a contemporary review of this field). Tensor network contractions of path integrals, rather than states, have also been introduced and successfully applied to these models, using the TRG (tensor renormalization group)  toolbox \cite{meurice_tensor_2020}. However, as the fundamental field theories of particle physics are continuous, it is also interesting to develop  tensor network formulations enabling their study directly, without applying any lattice discretization schemes.

Continuous tensor network states (cTNs) were first introduced in \cite{PhysRevLett.104.190405}, in the form of cMPS (continuous Matrix Product States). This formalism was further developed in later works \cite{haegeman_calculus_2013}, including higher dimensional extensions \cite{Jennings_2015,PhysRevX.9.021040} and even used as a numerical ansatz \cite{kara_gaussian_2021}. However, as these constructions rely on Fock space concepts, they are seemingly less equipped to deal with relativistic problems.  Recently, aiming at extending the cMPS to a relativistic setting, the one dimensional construction of RcMPS (relativistic cMPS) was introduced \cite{tilloy2021relativistic}. It uses a different operator basis, with respect to the particle creation and annihilation operators, which is more adequate for dealing with relativistic problems. In other tensor network techniques involving continuous fields, one may use continuous fields to contract lattice tensor networks for spin models \cite{PhysRevB.103.155130}. Finally, another type of tensor network states, MERA (multiscale entanglement renormalization ansatz) \cite{vidal_class_2008} has its continuum version - cMERA - too  \cite{PhysRevLett.110.100402,Nozaki:2012zj}. But in general, the field of continuous tensor network states is still far from being fully studied and understood, in spite of its obvious importance and relevance, and hence provides an interesting and challenging scientific playground.

		In this work,  we present a continuous, field-theoretic version of PEPS which is built from quantum fields. We therefore refer to them as cPEPS - continuous PEPS. The reason this construction sheds new light on continuous tensor-networks is twofold. To begin with, these states are well defined in any dimension and they are well suited to deal with relativistic field theories simply from the way they are formalized. Moreover, they inherit many of the appealing attributes of the PEPS such their ability to account for symmetries. Second, the cPEPS are formalized in a field-theoretic way. This simple fact is crucial as it brings this particular manifestation of a continuous tensor-network closer in spirit to relativistic quantum field theory, which has its own wide array of computational techniques \cite{srednicki_2007,Weinberg:1995mt} - specifically also of quantum informational calculations (see \cite{RevModPhys.90.035007} and references therein). Thus, it may hopefully serve as a bridge for implementing tensor-network techniques in quantum field theories and vice versa.
		
In section II, we present the cPEPS construction in detail and it is rigorously demonstrated how the ground-state of a free scalar field (Klein-Gordon) theory can be obtained from it, both exactly (with unlimited computational resources) or approximately. We study the accuracy of the approximation, show that it depends on the choice of a regularization scheme and propose a universal (regularization-independent) approach for tackling this problem. In section III, we present a bottom-up construction of the cPEPS as the continuum limit of a conventional, lattice PEPS construction, and explain how a cPEPS with continuous translational and rotational symmetry can be built. Finally, in section IV we explain in detail how states that are invariant under a general global symmetry can be directly constructed.
		
	In what follows we adhere to the common choice of dimensions $\hbar=c=1$. 
	The Einstein summation convention where repeated indices are being summed over is  assumed.

	\section{cPEPS}

We will begin our discussion of cPEPS by showing its construction for, perhaps, the simplest possible physical case, of a real scalar field in a $d+1$ dimensional flat (Minkowski) spacetime.
	 The Hilbert space of the field theory is constructed by choosing a foliation of space, and by defining the field and its respective conjugate momentum field over each slice (equal time surface), where the two satisfy the canonical commutation algebra. Dictated by the dynamics of the theory, there is then a unitary evolution from one slice to the other (time evolution). We choose the standard foliation in which spacetime is divided into constant $d$ dimensional (spatial) time slices.
	
	On each slice of constant time $t_{0}$ we introduce the  field operator $\phi\left(\textbf{x},t_{0}\right)$, and its conjugate momentum operator $\Pi\left(\textbf{x},t_{0}\right)$, satisfying  the canonical commutation relation
	\begin{equation}\label{canrel}
	 \left[\phi\left(\textbf{x},t_{0}\right),\pi\left(\textbf{y},t_{0}\right)\right]=i\delta^{\left(d\right)}\left(\textbf{x}-\textbf{y}\right).
	 \end{equation}

	Choosing an arbitrary time-slice (that is, fixing the time) we denote by 
$\left|\left\lbrace \phi\left(\textbf{x}\right)\right\rbrace\right\rangle$ a field configuration state:
\begin{equation}
	\left|\left\lbrace \phi\left(\textbf{x}\right)\right\rbrace\right \rangle=\underset{\mathbf{x}}{\otimes}\left|\phi\left(\textbf{x}\right)\right\rangle
\end{equation}
- that is, a product of eigenstates of the field operator everywhere. The field configuration states form a basis, in which one can span any field state. In particular, we choose to express our cPEPS in this basis and define it as
\begin{equation}\label{2.1}
	\left|\psi\right\rangle =\int\mathcal{D}\phi\left(\textbf{x}\right)\int\mathcal{D}v_{\alpha}\left(\textbf{x}\right)e^{\int d^{d}x\,\mathcal{A}\left[v_{\alpha},\nabla v_{\alpha},\phi,\nabla\phi\right]}\left|\left\lbrace \phi\left(\textbf{x}\right)\right\rbrace\right \rangle
\end{equation}
where
		\begin{equation}
			\mathcal{A}=-\frac{1}{2}Z_{\alpha\beta}\nabla v_{\alpha}\nabla v_{\beta}+V\left[v_{\alpha},\nabla v_{\alpha},\phi,\nabla\phi\right].
		\end{equation}
	Here $\phi\left(x\right)$ is a real scalar field - the physical field of interest. Additionally, $\left\{ v_{\alpha}\left(\textbf{x}\right)\right\} _{\alpha=1}^{D}$ is a set of $D$  fields which are called henceforth virtual fields and are being integrated over (their role is to allow for a locally contracted state of the physical field, in full analogy with lattice PEPS contractions, as we shall later show). Like $\phi$, the virtual fields are also real scalars. $Z_{\alpha\beta}$ is a $D\times D$ matrix whose eigenvalues have a nonnegative real part  and $V$ is a general analytic functional of the fields and their first derivatives. Following previous papers \cite{PhysRevLett.104.190405}-\cite{tilloy2021relativistic}, we shall call $D$ the (generalized) bond dimension. 

Note that these states may be expressed as the CTNS (continuous tensor network states) introduced in \cite{PhysRevX.9.021040}, and vice versa (see appendix \ref{appendixctns} for further details on the equivalence).

	To demonstrate what could be done with such states, we will restrict ourselves now to a case which can be handled analytically: Gaussian states, in which
 $\mathcal{A}$ is quadratic in both the physical and virtual fields and their first derivatives (other than being  analytically tractable, Gaussian states are interesting in their own right as they serve as the ground-states of non-interacting theories). The most general $\mathcal{A}$ of such a state may be written as follows:

	\begin{equation}\label{2.2}
		\begin{aligned}
			\mathcal{A}=&-\frac{1}{2}Z_{\alpha\beta}\nabla v_{\alpha}\left(\textbf{x}\right)\nabla v_{\beta}\left(\textbf{x}\right)-\frac{1}{2}A_{\alpha\beta}v_{\alpha}\left(\textbf{x}\right)v_{\beta}\left(\textbf{x}\right) \\
			&+z_{\alpha}\nabla v_{\alpha}\left(\textbf{x}\right)\nabla\phi\left(\textbf{x}\right)+a_{\alpha}v_{\alpha}\left(\textbf{x}\right)\phi\left(\textbf{x}\right)-\frac{c}{2}\phi^{2}\left(\textbf{x}\right)
		\end{aligned}
	\end{equation}

	Although one could, in principle, choose $\left\{ Z_{\alpha\beta},A_{\alpha\beta},z_{\alpha},a_{\alpha},c\right\}$ to be position dependent, we did not do this, and thus the above expression is translationally invariant. Thus, it 	
	 is natural from this point onward to work in Fourier (momentum) space. using the Fourier transform
	 \begin{equation}
	  \phi\left(\textbf{k}\right)=\int d^{d}k\;e^{-i\textbf{k}\cdot\textbf{x}}\phi\left(\textbf{x}\right).
	  \end{equation}
  (and similarly for the virtual fields) we obtain:
	\begin{equation}\label{momentum}
		\begin{aligned}
			&\left|\psi\right\rangle =\int\mathcal{D}\phi\left(\textbf{k}\right)\mathcal{D}v_{\alpha}\left(\textbf{k}\right)
			e^{\int\frac{d^{d}k}{\left(2\pi\right)^{d}}\mathcal{A}\left[v_{\alpha}\left(\textbf{k}\right),\phi\left(\textbf{k}\right)\right]}\left|\left\{ \phi\left(\textbf{k}\right)\right\} \right\rangle \\
			\mathcal{A}=&-\frac{1}{2}\left(A_{\alpha\beta}+Z_{\alpha\beta}k^{2}\right)\bar{v}_{\alpha}\left(\textbf{k}\right)v_{\beta}\left(\textbf{k}\right)-\frac{c}{2}\bar{\phi}\left(\textbf{k}\right)\phi\left(\textbf{k}\right) \\
			&+\frac{1}{2}\left(a_{\alpha}+z_{\alpha}k^{2}\right)\left(\bar{v}_{\alpha}\left(\textbf{k}\right)\phi\left(\textbf{k}\right)+\bar{\phi}\left(\textbf{k}\right)v_{\alpha}\left(\textbf{k}\right)\right).
		\end{aligned}
	\end{equation}
	Since these are real fields, they additionally satisfy $\bar{\phi}\left(\textbf{k}\right)=\phi\left(-\textbf{k}\right)$, even though it shall not be of importance for this paper. After integrating out the virtual fields we will end with a Gaussian state $\left|\psi_{D}\right\rangle$, quadratic in $\phi$:
	
	\begin{equation}\label{2.3}
		\left|\psi_{D}\right\rangle =\int\mathcal{D}\phi e^{-\frac{1}{2}\int\frac{d^{d}k}{\left(2\pi\right)^{d}}\omega_{D}\left(k\right)\bar{\phi}\left(\textbf{k}\right)\phi\left(\textbf{k}\right)}\left|\left\{ \phi\left(\textbf{k}\right)\right\} \right\rangle
	\end{equation}
Where we assume that $a_{\alpha}$, $z_{\alpha}$ are real; otherwise the generalization is straightforward):
		\begin{equation}
			\omega_{D}\left(k\right)=c+\frac{1}{2}\left(a_{\alpha}+z_{\alpha}k^{2}\right)\left(A+Zk^{2}\right)_{\alpha\beta}^{-1}\left(a_{\beta}+z_{\beta}k^{2}\right)
		\end{equation}
		
		\subsection{Approximation of the Free Vacuum}
		
		A more explicit expression for $\omega_{D}\left(k\right)$ can be put forth by using the matrix inversion formula:
				\begin{equation}
			\left(A+Zk^{2}\right)_{\alpha\beta}^{-1}=\frac{\text{adj}\left(A+Zk^{2}\right)_{\alpha\beta}}{\text{det}\left(A+Zk^{2}\right)}
		\end{equation}
		where $\text{adj}\left(M\right)$ is the adjugate matrix of $M$. From that we can tell that $\omega_{D}\left(k\right)$ is a rational function in the argument $k^{2}$, of order $D$ over $D$. The parameters which appear in the Gaussian cPEPS (\ref{2.1}) will be collectively denoted by $P=\left\{ Z_{\alpha\beta},A_{\alpha\beta},z_{\alpha},a_{\alpha},c\right\}$. We thus write
		\begin{equation}\label{2.4}
			\omega_{D}\left(k\right)=\frac{p_{0}\left(P\right)+p_{1}\left(P\right)k^{2}+...+p_{D}\left(P\right)k^{2D}}{1+q_{1}\left(P\right)k^{2}+...+q_{D}\left(P\right)k^{2D}}
		\end{equation}

	We have used the freedom to set one of the parameters of the rational function to 1. $p_{\alpha}\left(P\right),q_{\alpha}\left(P\right)$ are non-linear maps which are non-injective; note that while $2D-1$	parameters uniquely determine $\omega_{D}$, the number of parameters in the set $P$ is of order $D^{2}$. This redundancy can be eliminated by "gauge-fixing" conditions, common for tensor networks which are usually redundant constructions \cite{cirac2021matrix}, as addressed in previous continuum works \cite{PhysRevX.9.021040,kara_gaussian_2021,tilloy2021relativistic}. Yet this will not be necessary for our purposes.
	
	With (\ref{2.3}) being a Gaussian state, it is natural to inspect how well it approximates the ground-state (vacuum) of a free field theory, which is given by the Hamiltonian (see appendix B for more details):
		
		\begin{equation}\label{Hf}
			H_{f}=\frac{1}{2}\int\frac{d^{d}k}{\left(2\pi\right)^{d}}\left(\bar{\Pi}\left(\textbf{k}\right)\Pi\left(\textbf{k}\right)+\omega^2_{f}\left(k\right)\bar{\phi}\left(\textbf{k}\right)\phi\left(\textbf{k}\right)\right),
		\end{equation}
with the relativistic dispersion relation: \begin{equation}
	\omega^{2}_{f}\left(k\right)=m^{2}+k^{2}.
\end{equation}
 The free vacuum $\left|\Omega_{f}\right\rangle$ is given by:
		\begin{equation}\label{vac}
			\left|\Omega_{f}\right\rangle=\int\mathcal{D}\phi\left(\textbf{k}\right)e^{-\frac{1}{2}
				\int \frac{d^dk}{\left(2\pi\right)^d}\omega_{f}\left(k\right)\bar{\phi}\left(\textbf{k}\right)\phi\left(\textbf{k}\right)}\left|\left\{\phi\left(\textbf{k}\right)\right\} \right\rangle
		\end{equation}
	
		Given the functional form of $\omega_{D}\left(k\right)$, the approximation to some desired dispersion relation such as $\omega_{f}$ is an approximation by a rational function, known as the Padé approximant of a function \cite{baker_graves-morris_1996}.
	There are infinitely many different approximations which converge to $\omega_{f}\left(k\right)$ as $D\rightarrow\infty$. This fact can be seen, for instance, by noting that the Padé approximant, like the Taylor series it is based on, is localized around a particular base point of momentum $k_0$, and the choice of such base point is of course arbitrary. Nevertheless, there is both a theoretical and effective difference between these distinct	approximations, which will be addressed in the following subsection.\\
	
	It is of importance in rigorously demonstrating how the dispersion relation of a free relativistic field theory can be obtained from the Gaussian cPEPS~(\ref{2.1}). In appendix A we concretely construct such a state with mass $m$ using the continued fraction representation of the square root:
	\begin{equation}
		\sqrt{m^{2}+k^{2}}=m+\cfrac{k^{2}}{2m+\cfrac{k^{2}}{2m+\frac{k^{2}}{\ddots}}},
	\end{equation}
which is satisfied by choosing
\begin{equation}\label{params}
	\begin{aligned}
	&A_{\alpha,\beta}=\begin{cases} 	2m\delta_{\alpha,\beta} & \alpha \space \text{ even} \\
		0 & \alpha \space \text{ odd} \end{cases},\\
	&Z_{\alpha,\beta}=\frac{1}{2}\left(\delta_{\alpha,\beta-1}+\delta_{\alpha,\beta+1}\right)+
	\begin{cases} 	2m\delta_{\alpha,\beta} & \alpha \space \text{ odd} \\
		0 & \alpha \space \text{ even} \end{cases}
,\\
	&a_{\alpha}=0, \quad z_{\alpha}=\sqrt{2}\delta_{\alpha,1}, \quad c=m
	\end{aligned}
\end{equation}
for an even $D \rightarrow \infty$.

Note that in the numerical-variational approach to such Gaussian states, described in \cite{kara_gaussian_2021}, the $z_{\alpha}$ coupling between the gradients of virtual and physical fields was not required, while in the approach presented here, which is different and rather analytical, it is required.
	
	We see that when $D$ is infinite, the ground-state can be constructed as a cPEPS, exactly: the continued fraction is known to converge to $\omega_{f}\left(k\right)$ as $D\rightarrow\infty$. However, what if the continued fraction is truncated after $D$ steps?

	\subsection{Quantum Fidelity and Universality}
	
	We would like to inquire how well the Gaussian cPEPS~(\ref{2.2}) with a finite $D$ approximates the free vacuum $\left|\Omega_{f}\right\rangle$ (which is also a Gaussian cPEPS, but with an infinite $D$).
	A natural probe that we explore for that task is the quantum fidelity, which is defined for two pure states as:
		\begin{equation}
		F\left(\Omega_{f},\psi_{D};\Lambda\right)=\left|\left\langle \psi_{D}\middle|\Omega_{f}\right\rangle \right|
	\end{equation}
	
	To compute this quantity, we have chosen a Wilsonian regularization where the momentum modes are truncated at a cutoff $k=\Lambda$. The fidelity depends also on the cutoff $\Lambda$. The two states are assumed to be normalized, which bounds the fidelity from above and below: $0\leq F\left(\Omega_{f},\psi_{D};\Lambda\right)\leq1$.

	The fidelity takes maximal value, $1$, only when the two states $\left|\psi_{D}\right\rangle$ and $\left|\Omega_{f}\right\rangle$ are exactly identical (and hence the two functions $\omega_{D}\left(k\right)$ and $\omega_{f}\left(k\right)$). Since the fidelity is bounded, any difference  between the two functions must result in a smaller value for the fidelity. Assuming the fidelity is also a smooth functional of $\omega_{D}\left(k\right)$ and $\omega_{f}\left(k\right)$ (which holds true in this particular case, as to be shown below), it will increase as the two functions $\omega_{D}\left(k\right)$ and $\omega_{f}\left(k\right)$ get closer to each other within the interval $0\leq k\leq\Lambda$. Hence we can choose $\omega_{D}\left(k\right)$ to be the Padé approximant of $\omega_{f}\left(k\right)$ around some chosen base point $k_0$. The approximant diverges from the original function for large values of $k\gg k_0$ in this case. This can be seen as $\omega_{f}\left(k\right)$ asymptotically behaves as $k$, a behavior which a rational function in $k^2$ can ever achieve. On the interval $0\leq k\leq\Lambda$ the convergence to $\omega_{f}\left(k\right)$ is uniform. Although since the approximation is local, the rate of convergence does not depend on $\Lambda$. In conclusion, the fidelity increases when $D$ is increased while $\Lambda$ is kept fixed, and decreases when $D$ is kept fixed while $\Lambda$ is increased.

	This may be alarming since it seems that as the cutoff grows, a larger and larger bond dimension will be required to sustain a reasonable overlap between the two states. Furthermore, it implies that the bond dimension has no physical significance as it depends on the way the theory is regularized.
	However, in case we allow the parameters $P$ of the approximation to depend upon the cutoff themselves, it allows for rate of convergence to depend on $\Lambda$ as well. In which case, it enables, as the cutoff is sent to infinity, to tune the parameters in a way that makes the fidelity decay much less quickly, or perhaps even saturate to a non-zero value. 
	
	Given the Gaussianity of the particular case of study, the fidelity can be explicitly calculated (and even later expanded around the large cutoff limit). The cPEPS $\left|\psi_{D}\right\rangle$ and the free ground-state $\left|\Omega_{f}\right\rangle$
	both factorize into tensor products over momentum modes of harmonic oscillator ground-states:
	\begin{widetext}
		\begin{equation}
		\begin{aligned}
			&\left|\Omega_{f}\right\rangle =\underset{\textbf{k}}{\overset{\Lambda}{\prod}}\left|\Omega_{f}\left(\textbf{k}\right)\right\rangle =\underset{\textbf{k}}{\overset{\Lambda}{\prod}}\left(\frac{\omega_{f}\left(k\right)}{\pi}\right)^{1/4}
			\int d\phi\left(\mathbf{k}\right)
			e^{-\frac{1}{2}
				\omega_{f}\left(k\right)\bar{\phi}\left(\textbf{k}\right)\phi\left(\textbf{k}\right)}\left|\phi\left(\textbf{k}\right)\right\rangle \\
			&\left|\psi_{D}\right\rangle =\underset{\textbf{k}}{\overset{\Lambda}{\prod}}\left|\psi_{D}\left(\textbf{k}\right)\right\rangle =\underset{\textbf{k}}{\overset{\Lambda}{\prod}}\left(\frac{\omega_{D}\left(k\right)}{\pi}\right)^{1/4}
						\int d\phi\left(\mathbf{k}\right)
						e^{-\frac{1}{2}\omega_{D}\left(k\right)\bar{\phi}\left(\textbf{k}\right)\phi\left(\textbf{k}\right)}\left|\phi\left(\textbf{k}\right)\right\rangle \\
		\end{aligned}
	\end{equation}
	The fidelity therefore factorizes as well:
	\begin{align}
	\left\langle \psi_{D}\left(\textbf{k}\right)\middle|\Omega_{f}\left(\textbf{k}\right)\right\rangle= 
\left(\frac{\omega_{f}\left(k\right)\omega_{D}\left(k\right)}{\pi^{2}}\right)^{\frac{1}{4}}\int d\phi\left(\textbf{k}\right)e^{-\frac{1}{2}\left(\omega_{f}\left(k\right)+\omega_{D}\left(k\right)\right)\bar{\phi}\left(\textbf{k}\right)\phi\left(\textbf{k}\right)}= 
\left(\frac{\omega_{f}\left(k\right)\omega_{D}\left(k\right)}{\pi^{2}}\right)^{\frac{1}{4}}\left(\frac{2\pi}{\omega_{f}\left(k\right)+\omega_{D}\left(k\right)}\right)^{\frac{1}{2}}
	\end{align}
	We finally obtain
	\begin{equation}\label{fid}
		F\left(\Omega_{f},\psi_{D};\Lambda\right)=\underset{\textbf{k}}{\overset{\Lambda}{\prod}}\left|\left\langle \psi_{D}\left(\textbf{k}\right)\middle|\Omega_{f}\left(\textbf{k}\right)\right\rangle \right|= \underset{\textbf{k}}{\overset{\Lambda}{\prod}}\left(\frac{2\sqrt{\left|\omega_{D}\left(k\right)\right|\omega_{f}\left(k\right)}}{\left|\omega_{D}\left(k\right)+\omega_{f}\left(k\right)\right|}\right)^{\frac{1}{2}}
	\end{equation}
\end{widetext}

	In order to deal with the product inside (\ref{fid}) in the continuum limit $\Lambda\rightarrow\infty$, we need to be more precise with the definition of the Wilsonian regularization. We begin from regularizing the field theory by placing it on a lattice with $N$ sites and spacing $\frac{1}{\Lambda}$. The volume $V=\frac{N}{\Lambda^{d}}$ of the $d$-dimensional spatial space is taken to be finite yet arbitrarily large and likewise the cutoff $\Lambda$ is kept fixed. Consequently, upon taking the logarithm the sum can be turned into an integral:
	
	\begin{equation}
		\log F=\frac{V}{2}\int_{\Omega_{\Lambda}}\frac{d^{d}k}{\left(2\pi\right)^{d}}\log\left(\frac{2\sqrt{\left|\omega_{D}\left(k\right)\right|\omega_{f}\left(k\right)}}{\left|\omega_{D}\left(k\right)+\omega_{f}\left(k\right)\right|}\right)
	\end{equation}
	
	Where $\int_{\Omega_{\Lambda}}$ is an integral in the domain $0\leq k\leq\Lambda$. The scaling of the problem becomes more transparent by using the dimensionless variable: $\bar{\text{\textbf{k}}}=\frac{\textbf{k}}{\Lambda}$:
	
	\begin{equation}
		\log F=\frac{N}{2}\int_{\Omega_{1}}\frac{d^{d}\bar{k}}{\left(2\pi\right)^{d}}\log\left(\frac{2\sqrt{\left|\omega_{D}\left(\bar{k}\right)\right|\omega_{f}\left(\bar{k}\right)}}{\left|\omega_{D}\left(\bar{k}\right)+\omega_{f}\left(\bar{k}\right)\right|}\right)
	\end{equation}
where $\Omega_{1}$ is the unit sphere.
		It is then seen that the fidelity decays exponentially with the system's size (as what multiplies $N$ must be a negative number since each harmonic ground-state of a single mode $\textbf{k}$ was taken to be normalized).
	    Such behavior between two distinguished states of a quantum many-body system is known to hold in many examples (e.g. \cite{PhysRevLett.106.055701,doi:10.1142/S0217979210056335})  and in particular has been shown to also be generic for Matrix Product States (MPS) \cite{PhysRevB.76.104420} (other possibilities may also occur \cite{PhysRevB.79.092405},\cite{doi:10.1142/S0217979210056335}). In principle, the fidelity would go to zero as $\Lambda\rightarrow\infty$ or ${V\rightarrow\infty}$ even if the two dispersion relations are infinitesimally close to each other but not exactly equal (this phenomena is also known as Anderson's orthogonality catastrophe \cite{PhysRevLett.18.1049}). In which case, we see that a relatable quantity which bears its own significance exists, this is the fidelity per-site, defined by:
		
		\begin{equation}
			\mathcal{F}=\underset{N\rightarrow\infty}{\lim}F^{\frac{1}{N}}
		\end{equation}
			This may be defined alternatively in terms of the logarithm: $\log\,\mathcal{F}\left(\Omega_{f},\psi_{D};\Lambda\right)=\underset{N\rightarrow\infty}{\lim}\frac{1}{N} \log\,F\left(\Omega_{f},\psi_{D};\Lambda\right)$. \\
		
		The fidelity per-site $\mathcal{F}$ is well defined in the continuum limit $\Lambda\rightarrow\infty,V\rightarrow\infty$ as it has no dependence on the volume or the cutoff - unless inserted explicitly through $\omega_{D}$ through the parameters $P$. We now explore this option, and for that we shall rescale the dispersion relations $\omega_{D}\left(k\right)$, $\omega_{f}\left(k\right)$ in the dimensions of the cutoff,  similarly to the definition of $\bar{k}$. Hence we define:
	\begin{equation}
		\begin{aligned}
			&\omega_{f}\left(\bar{k}\right)=\Lambda\tilde{\omega}_{f}\left(\bar{k}\right) \\
			&\omega_{D}\left(\bar{k}\right)=\Lambda\tilde{\omega}_{D}\left(\bar{k}\right)
		\end{aligned}
	\end{equation}
		We see that we may expand $\tilde{\omega}_{f}\left(k\right)$ around $\Lambda\rightarrow\infty$:
	\begin{equation}\label{2.7}
		 \tilde{\omega}_{f}\left(\bar{k}\right)=\sqrt{\bar{k}^{2}+\left(\frac{m}{\Lambda}\right)^{2}}=\bar{k}+\frac{1}{2\bar{k}^{2}}\left(\frac{m}{\Lambda}\right)^{2}+O\left(\left(\frac{m}{\Lambda}\right)^{4}\right)
	\end{equation}
and	
	\begin{equation}\label{2.8}
		\tilde{\omega}_{D}\left(\bar{k}\right)=\frac{\tilde{p}_{0}\left(P\right)+\tilde{p}_{1}\left(P\right)\bar{k}^{2}+..\tilde{p}_{D}\left(P\right)\bar{k}^{2D}}{1+\tilde{q}_{1}\left(P\right)\bar{k}^{2}+..\tilde{q}_{D}\left(P\right)\bar{k}^{2D}}
	\end{equation}
	
	The rescaled parameters can be related back to the original ones by:
	
	\begin{equation}
		\begin{aligned}
			&p_{\alpha}\left(P\right)=\Lambda^{1-2\alpha}\tilde{p}_{\alpha}\left(P\right) \\
			&q_{\alpha}\left(P\right)=\Lambda^{-2\alpha}\tilde{q}_{\alpha}\left(P\right)
		\end{aligned}
	\end{equation}

	We note that this is an arbitrary choice, since the extra power of $\Lambda$ could have been otherwise extracted from the denominator.
		Similarly to (\ref{2.7}), an expansion around $\Lambda\rightarrow\infty$ for  $\tilde{\omega}_{D}\left(\bar{k}\right)$ can be made as well. The form of this expansion will depend of course on the way we have chosen the parameters $P$ to depend on the cutoff $\Lambda$. We therefore chose $\tilde{p}_{\alpha},\tilde{q}_{\alpha}$ to be separated very generally into a cutoff independent term and an irrelevant term:
	\begin{equation}
		\begin{aligned}
			&\tilde{p}_{\alpha}\left(P\right)=\tilde{p}_{\alpha}^{\left(0\right)}\left(P\right)+\tilde{p}_{\alpha}^{\left(irr\right)}\left(P;\Lambda\right) \\
			&\tilde{q}_{\alpha}\left(P\right)=\tilde{q}_{\alpha}^{\left(0\right)}\left(P\right)+\tilde{q}_{\alpha}^{\left(irr\right)}\left(P;\Lambda\right)
		\end{aligned}
	\end{equation}

	$\tilde{p}_{\alpha}^{\left(irr\right)},\tilde{q}_{\alpha}^{\left(irr\right)}$ are defined to vanish as $\Lambda\rightarrow\infty$. This definition is necessary since the fidelity is bounded for normalized states, therefore the parameters $P$ cannot be chosen in such a way that the fidelity diverges.
	
Since the decay of $\tilde{\omega}_{f}\left(\bar{k}\right)$ in $\Lambda$ is polynomial, it is an evident choice to let $\tilde{\omega}_{D}\left(\bar{k}\right)$ decay polynomially as well, thus an expansion in powers of $\frac{1}{\Lambda}$ for $\tilde{\omega}_{D}\left(\bar{k}\right)$ is possible as well, similarly to (\ref{2.7}). However, the decay can be in principle of a different class instead, such as an exponential one. 
	
	Derived from this assumption, $\tilde{\omega}_{D}\left(\bar{k}\right)$ may be expanded in the same way around $\Lambda\rightarrow\infty$:
	
	\begin{equation}
		\tilde{\omega}_{D}\left(\bar{k}\right)=\tilde{\omega}_{D}^{\left(0\right)}\left(\bar{k}\right)+\tilde{\omega}_{D}^{\left(irr\right)}\left(\bar{k};\Lambda\right)
	\end{equation}
where
	\begin{equation}
		\tilde{\omega}_{D}^{\left(0\right)}\left(\bar{k}\right)=\frac{\tilde{p}_{0}^{\left(0\right)}\left(P\right)+\tilde{p}_{1}^{\left(0\right)}\left(P\right)\bar{k}^{2}+..\tilde{p}_{D}^{\left(0\right)}\left(P\right)\bar{k}^{2D}}{1+\tilde{q}_{1}^{\left(0\right)}\left(P\right)\bar{k}^{2}+..\tilde{q}_{D}^{\left(0\right)}\left(P\right)\bar{k}^{2D}}
	\end{equation}
	
	To conclude, putting together (\ref{2.7}) and (\ref{2.8}) yields:
	\begin{equation}
		\log\,\mathcal{F}=\frac{1}{2}\int_{\Omega_{1}}\frac{d^{d}\bar{k}}{\left(2\pi\right)^{d}}\log\left(\frac{2\sqrt{\left|\tilde{\omega}_{D}^{\left(0\right)}\left(\bar{k}\right)\right|\bar{k}}}{\left|\tilde{\omega}_{D}^{\left(0\right)}\left(\bar{k}\right)+\bar{k}\right|}\right)+\mathcal{F}^{\left(irr\right)}\left(\Lambda\right)
	\end{equation}
		The first term has no dependence in $\Lambda$ at all, and $\mathcal{F}^{\left(irr\right)}\left(\Lambda\right)$ is a function
		that vanishes as $\Lambda\rightarrow\infty$.
		
		In summary, we deduce that while in this particular example the fidelity itself is doomed to vanish in the continuum limit, there exist a quantity with a universal term that can in principle be variationally optimized.
	
	\section{cPEPS From PEPS}
	
	\subsection{PEPS With Fields}
	
	Starting from the lattice, we use the PEPS formalism to construct
	a state whose continuum limit is the cPEPS (\ref{2.1})
	or a generalization thereof. We will first show how to do it for the simple case of a single scalar field, and then generalize to more possibilities, including scenarios with global symmetries.
	
	On a constant time slice where the Hilbert space is defined, we consider a spatial lattice $\mathbb{Z}^d$, whose sites  are labeled by vectors of the form $\mathbf{x} \in \epsilon\mathbb{Z}^d \equiv \mathcal{L}$, where   $\epsilon > 0$ is the lattice spacing. We use $\left\{ \hat{\mathbf{e}}_{i}\right\} _{i}^{d}$ to denote the unit vectors in all the positive directions. Thus, for example, the sites $\mathbf{x}$ and $\mathbf{x}+\epsilon\hat{\mathbf{e}}_i$ are nearest neighbors in the ${1\leq i\leq d}$ direction, separated by a single unit of lattice spacing, $\epsilon$.
	
	We would like to construct a PEPS for the state of a lattice scalar field $\phi\left(\mathbf{x}\right)$: on each lattice site $\mathbf{x}$, we introduce a physical Hilbert space, $\mathcal{H}_{\text{phys}}\left(\mathbf{x}\right)$, spanned by eigenstates of the on-site field operator $\phi\left(\mathbf{x}\right)$. This can be either a real scalar field (as before) or a complex one (corresponding, as usual, to a pair of real scalar fields). Thus, a good basis for the local physical Hilbert space is $\left|\phi\left(\mathbf{x}\right)\right\rangle$ in the real case and $\left|\phi\left(\mathbf{x}\right),\overline{\phi}\left(\mathbf{x}\right)\right\rangle$ in the complex one (where $\overline{z}$ represents complex conjugation). Below, we will adapt the notation of the complex case, and the real case will follow from it in a straightforward manner. 
	
	The physical Hilbert space of the whole system is 
	\begin{equation}
		\mathcal{H}_{\text{phys}}=\underset{\mathbf{x}\in \mathcal{L}}{\bigotimes}\mathcal{H}_{\text{phys}}\left(\mathbf{x}\right)
		\end{equation}
	However, as usual in the construction of PEPS, the physical degrees of freedom are not sufficient: to contract the local ingredients together, we have to introduce auxiliary or \emph{virtual} fields to our system. That is, on each lattice site $\mathbf{x} \in \mathcal{L}$, we introduce $2d$ extra Hilbert spaces, associated with the legs going out of and into it. On each outgoing leg, in direction $i$, we define the virtual fields $\left\{\chi_i^{\alpha}\left(\mathbf{x}\right)\right\}_{\alpha=1}^D$, while on each ingoing leg we introduce similar fields $\left\{\eta_i^{\alpha}\left(\mathbf{x}\right)\right\}_{\alpha=1}^D$. All virtual fields are either real or complex scalars, depending on the nature of the physical field $\phi\left(\mathbf{x}\right)$. As in the physical case, we will generally treat them as complex below. The number of fields we introduce on each leg, $D$, is nothing but the bond dimension mentioned above; we are free to choose any $D \geq 1$ that we want to build our PEPS - increasing $D$ will allow our PEPS to depend on more parameters - which implies more freedom for the optimization problem whose solution is sought with the PEPS. Different physical scenarios will impose different constraints on the required minimal bond dimension, and we have seen an example for that in our previous construction. However, whatever $D$ is, we unite the Hilbert spaces of all the virtual fields of $\mathbf{x} \in \mathcal{L}$ to $\mathcal{H}_\text{virt}\left(\textbf{x}\right)$. We denote its configuration states by
			\begin{equation}
		\left|\bm{\chi}\left(\textbf{x}\right),\bm{\eta}\left(\textbf{x}\right)\right\rangle:=\underset{i=1}{\overset{d}{\bigotimes}}\left|\chi_{i}\left(\textbf{x}\right)\right\rangle\left|\eta_{i}\left(\textbf{x}\right)\right\rangle
	\end{equation}
(where the $\alpha$ indices were omitted for simplicity). In a sense, $\bm{\chi}\left(\textbf{x}\right)=\left(\chi_{1}\left(\textbf{x}\right),...,\chi_{d}\left(\textbf{x}\right)\right)$ and ${\bm{\eta}\left(\textbf{x}\right)=\left(\eta_{1}\left(\textbf{x}\right),...,\eta_{d}\left(\textbf{x}\right)\right)}$ can be thought of as spatial vector fields.

On each site $\mathbf{x}\in\mathcal{L}$ we can therefore define the local Hilbert space
				\begin{equation}
			\mathcal{H}\left(\textbf{x}\right)=\mathcal{H}_{\text{physical}}\left(\textbf{x}\right)\otimes\mathcal{H}_{\text{virt}}\left(\textbf{x}\right).
		\end{equation} 
	involving both the physical and virtual degrees of freedom.
		Any state $\left|A\left(\textbf{x}\right)\right\rangle \in \mathcal{H}\left(\textbf{x}\right)$  may be expanded as
		\begin{widetext}
			\begin{equation}\label{3.2}
				\left|A\left(\textbf{x}\right)\right\rangle =\int \underset{i}{\prod}\left( d\chi_{i}\left(\textbf{x}\right)d\eta_{i}\left(\textbf{x}\right)\right)d\phi\left(\textbf{x}\right)A\left( \phi\left(\textbf{x}\right),\bm{\chi}\left(\textbf{x}\right),\bm{\eta}\left(\textbf{x}\right)\right)
				\left|\phi\left(\textbf{x}\right)\right\rangle\otimes\left|\bm{\chi}\left(\textbf{x}\right),\bm{\eta}\left(\textbf{x}\right)\right\rangle
			\end{equation}
(where integration on complex conjugates and/or different $\alpha$ components of the virtual fields is implicitly assumed).

$\underset{\mathbf{x}\in\mathcal{L}}{\bigotimes}\left|A\left(\mathbf{x}\right)\right\rangle$ is not the state we need: it is both a product state, with no correlations among the lattice sites, and it also involves some virtual degrees of freedom which have nothing to do with the physical Hilbert space $\mathcal{H}_{\text{phys}}$. Both issues are addressed, as usual, by projecting the virtual degrees of freedom onto maximally entangled pair states, connecting the virtual fields of both sides of each link; in our case, we choose
	\begin{equation}\label{3.1}
			\left|L_{i}\left(\textbf{x}\right)\right\rangle=\int  d\chi_{i}\left(\textbf{x}\right)d\eta_{i}\left(\textbf{x}+\epsilon\hat{\mathbf{e}}_{i}\right) \delta\left(\chi_{i}\left(\textbf{x}\right)-\eta_{i}\left(\textbf{x}+\epsilon\hat{\mathbf{e}}_{i}\right)\right) \\
			\delta\left(\bar{\chi}_{i}\left(\textbf{x}\right)-\bar{\eta}_{i}\left(\textbf{x}+\epsilon\hat{\mathbf{e}}_{i}\right)\right) \left|\chi_{i}\left(\textbf{x}\right)\right\rangle \left|\eta_{i}\left(\textbf{x}+\epsilon\hat{\mathbf{e}}_{i}\right)\right\rangle
	\end{equation}
		\end{widetext}
	defined on each link. The extension to $D>1$ is straightforward, and in the real field case the second delta function(s) are removed.

\begin{figure}[h!]
		\includegraphics[width=0.9\columnwidth]{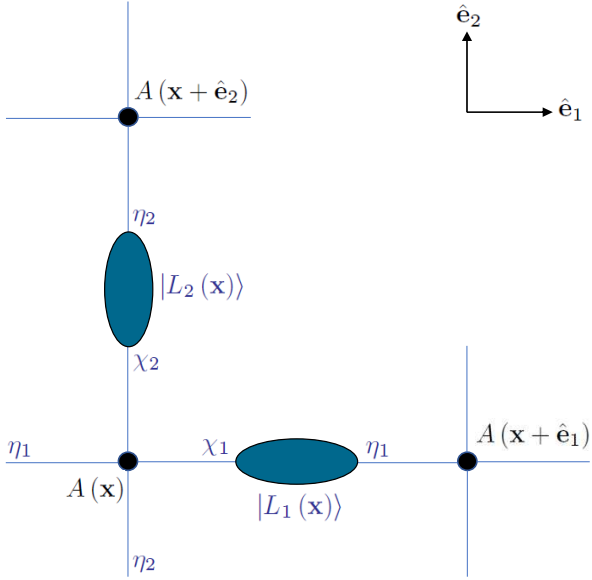}
		\caption{An example of a two-dimensional PEPS with fields. The lattice sites are marked with the black dots and the virtual fields with the ingoing and outgoing blue lines to each site. Two outgoing and ingoing virtual fields of two neighboring sites are contracted by the links $L_i$, which is illustrated by the oval connecting the two lines.}
		\label{Fig1}
	\end{figure}

	The PEPS is obtained by projecting $\underset{\mathbf{x}\in\mathcal{L}}{\bigotimes}\left|A\left(\mathbf{x}\right)\right\rangle$
	onto the maximally entangled states on all the links:
	\begin{equation}
	\left|\psi\right\rangle = \underset{\mathbf{x}\in\mathcal{L},i=1,...,d}{\bigotimes}	\left\langle L_i\left(\mathbf{x}\right)\right|
	\underset{\mathbf{x}\in\mathcal{L}}{\bigotimes}\left|A\left(\mathbf{x}\right)\right\rangle \in \mathcal{H}_{\text{phys}}
	\end{equation}
	A two-dimensional example for the construction detailed above is depicted in Fig. (\ref{Fig1}).
	
	We have not only eliminated the virtual degrees of freedom, we actually used them for inducing correlations among the lattice sites. Furthermore, this state will satisfy the area law, since for any bipartition we wish to make, we will get that the entanglement entropy comes from the maximally entangled states on the links which cross the boundary, all contributing the same.
	
\subsection{Spatial Symmetries}	
	We would like to require the PEPS $\left|\psi\right\rangle$ to be invariant under translations and under the transformations which preserve the lattice (its point group). This is a reasonable condition if we wish the state to become invariant under continuous translations and rotations $SO\left(d\right)$ in the continuum limit (the rest of the Poincaré group is explicitly broken by the choice of the quantization scheme). To that end we require the PEPS $\left|\psi\right\rangle$  to be invariant under discrete lattice translations and also under rotations preserving the lattice. As for the latter, where it is for the simple choice of a square lattice we will suffice to require invariance under rotations of $\frac{\pi}{2}$ in all possible planes (we do not make any prior demands regarding reflections and parity).
	
	The discrete translations $T\left(\boldsymbol{a}\right)$ for some displacement vector $\boldsymbol{a}$ on the lattice are defined to act in a very simple way on both the physical and virtual fields, 
	\begin{equation}\label{3.5}
		T\left(\boldsymbol{a}\right)\left|\phi\left(\textbf{x}\right)\right\rangle=\left|\phi\left(\textbf{x}+\boldsymbol{a}\right)\right\rangle 
	\end{equation}
It  acts exactly the same on the virtual
	fields $\chi_{i},\eta_{i}$.
	If not broken to begin with, the discrete translational symmetry on the lattice will be straightforwardly enhanced to a continuous translational symmetry in the continuum limit. Imposing translation symmetry on the PEPS is very simple, and can be done by requiring all the functions $A\left( \phi\left(\textbf{x}\right),\bm{\chi}\left(\textbf{x}\right),\bm{\eta}\left(\textbf{x}\right)\right)$ from  (\ref{3.2}) to be the same all around the lattice (independent of the coordinates). Similarly, we use the same $\left|L_i\left(\mathbf{x}\right)\right\rangle$ states on all the links in the same direction.
	
	Next we consider lattice rotations. Denote by $\Lambda_{ij}$ rotations of $\frac{\pi}{2}$ in the plane spanned by $\hat{\mathbf{e}}_{i}$ and $\hat{\mathbf{e}}_{j}$ (with no loss of generality, we assume that $i<j$). The coordinate rotation is given by
		\begin{equation}
		\Lambda_{ij}\textbf{x}=\Lambda_{ij}\left(x_{1},..,x_{i},..,x_{j},..x_{d}\right)=\left(x_{1},..,-x_{j},..,x_{i},..x_{d}\right)
	\end{equation}
	
	The physical field is scalar and thus its eigenstates transform as:
	\begin{equation}
		\left|\phi\left(\textbf{x}\right)\right\rangle \rightarrow \left|\phi\left(\Lambda_{ij}\textbf{x}\right)\right\rangle
	\end{equation}
	
	Nevertheless, the virtual fields must transform in a different manner:
	\begin{equation}\label{3.6}
		\begin{aligned}
			&\left|\chi_{i}\left(\textbf{x}\right)\right\rangle  \rightarrow\left|\chi_{j}\left(\Lambda_{ij}\textbf{x}\right)\right\rangle \\
			&\left|\chi_{j}\left(\textbf{x}\right)\right\rangle  \rightarrow\left|-\eta_{i}\left(\Lambda_{ij}\textbf{x}\right)\right\rangle \\
			&\left|\eta_{i}\left(\textbf{x}\right)\right\rangle  \rightarrow\left|\eta_{j}\left(\Lambda_{ij}\textbf{x}\right)\right\rangle \\
			&\left|\eta_{j}\left(\textbf{x}\right)\right\rangle  \rightarrow\left|-\chi_{i}\left(\Lambda_{ij}\textbf{x}\right)\right\rangle 
		\end{aligned}
	\end{equation}
	To explain why, we recall that these fields are associated with directions - they are components of spatial vector fields. Therefore, they must follow the rotation of links, giving rise to the above rotation rules. 
	
		Requiring the fiducial state $\left|A\left(\textbf{x}\right)\right\rangle$ to be invariant under (\ref{3.6}) is equivalent to requiring the wave-function in (\ref{3.2}), $A\left(\phi\left(\textbf{x}\right),\bm{\chi}\left(\textbf{x}\right),\bm{\eta}\left(\textbf{x}\right)\right)$ to be symmetric under the permutations of the virtual fields dictated by (\ref{3.6}) for all ${i<j}$:
		\begin{equation}\label{rot}
			\begin{aligned}
				A&\left(\phi,..,\chi_{i},..,\chi_{j},..\eta_{i},..,\eta_{j},..\right)\\
				=A&\left(\phi,..,\chi_{j},..,-\eta_{i},..\eta_{j},..,-\chi_{i},..\right)\\
				=A&\left(\phi,..,-\eta_{i},..,-\eta_{j},..-\chi_{i},..,-\chi_{j},..\right)\\
				=A&\left(\phi,..,-\eta_{j},..,\chi_{i},..-\chi_{j},..,\eta_{i},..\right)
			\end{aligned}
		\end{equation}
	One also has to make sure that the maximally entangled states on the links, $\left|L_i\left(\mathbf{x}\right)\right\rangle$ are rotated in an invariant way (in this case, we need to demand that in an ${i-j}$ rotation, the states belonging to these two directions are exchanged, and the others are left intact; overall, the product of all link states remains the same). This is satisfied by the states we picked for our demonstration above.

		In order to obtain the desired form of the cPEPS (\ref{2.1}) in the continuum limit, we write $A\left(\phi,\bm{\chi},\bm{\eta}\right)$ as an exponential:
		\begin{equation}\label{A}
		A\left(\phi,\bm{\chi},\bm{\eta}\right)=e^{\mathcal{A}\left(\phi,\bm{\chi},\bm{\eta}\right)}
	\end{equation}
Here, $\mathcal{A}$ respects the spatial symmetries (\ref{3.5}) and (\ref{rot}). We will now focus on a more specific ansatz,
\begin{equation}
		\mathcal{A}=K\left(\bm{\chi},\bm{\eta}\right)+V_{0}\left(\bm{\chi},\bm{\eta},\phi\right)
	\end{equation}
	where
	$V_{0}$ is a general analytic function of the fields that is required to be invariant under  (\ref{rot}). $K$ is defined by
	\begin{equation}
		K\left(\bm{\chi},\bm{\eta}\right)=-\frac{Z_{0}}{2}\underset{i}{\overset{d}{\sum}}\left|\chi_{i}\left(\textbf{x}\right)-\eta_{i}\left(\textbf{x}\right)\right|^{2} 
	\end{equation}

	\subsection{The Continuum Limit}
		The projection onto the maximally entangled link states  $\left|L_i\left(\mathbf{x}\right)\right\rangle$ allows us to use the delta functions, eliminate the $\eta_i$ Hilbert spaces and exchange any appearance of $\eta_{i}\left(\textbf{x}\right)$ by $\chi_{i}\left(\textbf{x}-\epsilon\hat{\mathbf{e}}_{i}\right)$. Collecting all of the above, the physical state may then be written as (here $D=1$ but the generalization to higher bond dimensions is straight forward):
		\begin{equation}\label{3.7}
			\left|\psi\right\rangle =  \int\underset{\textbf{x}}{\Pi}d\phi\left(\textbf{x}\right)\underset{\textbf{x},i}{\Pi}d\chi_{i}\left(\textbf{x}\right) 
			e^{\underset{\textbf{x}}{\sum}\mathcal{A}\left(\phi\left(\textbf{x}\right),\bm{\chi}\left(\textbf{x}\right)\right)} 
			\underset{\textbf{x}}{\Pi}\left|\phi\left(\textbf{x}\right)\right\rangle\left|\bm{\chi}\left(\textbf{x}\right)\right\rangle
		\end{equation}
		In the above expression appears the same function $\mathcal{A}$ as it was presented previously, only now with the insertions of $\eta_{i}\left(\textbf{x}\right)$ replaced with $\chi_{i}\left(\textbf{x}-\epsilon\hat{\mathbf{e}}_{i}\right)$ as explained.

	From this point the continuum limit is taken in the conventional way. That is, the dimensionless fields and parameters are exchanged by their continuum counterparts which obtain their mass dimension by an appropriate power of $\epsilon$.
	
		Namely, the fields $\phi,\chi_{i}$ which appear in (\ref{3.7}) are defined on the lattice and are dimensionless. For taking the continuum limit, we need to exchange summation by integration in the exponent of (\ref{3.7}). Since the integration measure is dimensionful, this requires the introduction of dimensionful fields to compensate for the originally dimensionless expression. $\left[\phi\right]$, the dimension of the physical field $\phi$ is chosen to have the standard dimension for the free relativistic field \footnote{This is set for instance from the (dimensionless) Klein-Gordon action: $S=\int d^{d+1}x\left(-\frac{1}{2}\partial^{\mu}\phi\partial_{\mu}-\frac{1}{2}m^2\phi^{2}\right)$ (the the kinetic term's coefficient is normalized to one by convention).}: $\left[\phi\right]=\frac{d-1}{2}$. This is a necessary choice as this degree of freedom also serves in a physical state such as the free vacuum (\ref{vac}).
		
		We define the "renormalized" physical field $\phi^{\left(R\right)}\left(\textbf{x}\right)=\epsilon^{-\left[\phi\right]}\phi\left(\textbf{x}\right)$.
		Since the virtual fields are eventually integrated out  there exists a freedom in choosing their dimension (see appendix \ref{appcfrac} for more details). We hence do not commit to a specific choice and denote their dimension by $\left[\chi\right]$. We  define $\chi_{i}^{\left(R\right)}\left(\textbf{x}\right)=\epsilon^{-\left[\chi\right]}\chi_{i}\left(\textbf{x}\right)$ accordingly. \\
	
	\begingroup
	Define $Z=\epsilon^{2-d+2\left[\chi\right]}Z_{0}$. We tune $Z_{0}$
	such that $Z$ will be finite in the continuum limit $\epsilon \rightarrow 0$. This immediately gives rise to the kinectic term
	\begin{equation}\label{kinetic}
		\underset{\textbf{x}}{\sum}K\left(\bm{\chi}\right)\rightarrow-\frac{1}{2}Z\int d^{d}x\underset{i=1}{\overset{d}{\sum}}\left|\nabla_{i}\chi^{R}_{i}\left(\textbf{x}\right)\right|^{2}
	\end{equation}
We further introduce the virtual field $v\left(\mathbf{x}\right)$,
\begin{equation}
	\chi_i^{(R)}\left(\mathbf{x}\right) \equiv v\left(\mathbf{x}+\frac{\epsilon}{2}\hat{\mathbf{e}}_i\right)
\end{equation}
which turns \ref{kinetic} into the stanard kinetic term for scalar fields. \endgroup
Finally, as was done in detail for $K$, the function $V_{0}\left(\chi_{i},\eta_{i},\phi\right)$ similarly becomes, by following all the steps described above, the functional  $V\left[v,\nabla v,\phi^{\left(R\right)}\right]$ (in fact, (\ref{2.1}) also depends on $\nabla\phi$, we will later show how this is achieved in the end of this section).
	
The mapping between the two can be defined in a formal way through the series expansion of $V$ in powers of the fields in their first derivatives and then identifying $v\left(\textbf{x}\right)$ with $\chi_{i}\left(\textbf{x}\right)$ and $\nabla_{i}v\left(\textbf{x}\right)$ with $\chi_{i}\left(\textbf{x}\right)-\eta_{i}\left(\textbf{x}\right)$.
For example, $V=a\left(\nabla v\right)^{2}v$ can be constructed from $V_{0}=a_{0}\left|\chi_{i}\left(\textbf{x}\right)-\eta_{i}\left(\textbf{x}\right)\right|^{2}\chi_{i}+\left(\text{permutations}\right)$. The permutations under (\ref{3.7}) ensure the rotational symmetry. Similarly to before, the connection between the parameters is given by  $a=\frac{1}{4}\epsilon^{2-d+3\left[\chi\right]}a_0$; the factor of $\frac{1}{4}$ comes to account for the 4 terms connected by the rotational transformation which collapse into a single one in the continuum limit.

Now that we are fully done, we can conveniently drop the $R$ superscript. Additional virtual fields can easily be added by following the same steps above, so we may consider a general bond dimension $D>1$. We then arrive to the continuum limit of (\ref{3.7}) which is the desired cPEPS:
		\begin{equation}\label{3.8}
		\begin{aligned}
			\left|\psi\right\rangle &=\int\mathcal{D}\phi\left(\textbf{x}\right)\mathcal{D}v_{\alpha}\left(\textbf{x}\right)
			e^{\int d^{d}x\mathcal{A}\left[v_{\alpha},\nabla v_{\alpha},\phi\right]}\left|\left\lbrace \phi\left(\textbf{x}\right)\right\rbrace\right\rangle \\
			&\mathcal{A}=-\frac{1}{2}Z_{\alpha\beta}\nabla_{i}\overline{v}_{\alpha}\nabla_{i} v_{\beta}+V\left[v_{\alpha},\nabla v_{\alpha},\phi\right]
		\end{aligned}
	\end{equation}

	\begin{figure}[h!]
		\includegraphics[width=0.9\columnwidth]{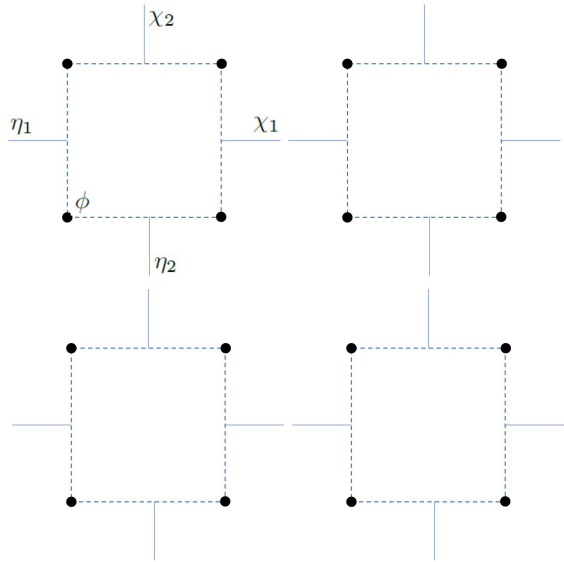}
		\caption{An example for a two-dimensional square lattice with a  $\mathcal{H}_{\text{phys}}$ of larger support. The block is of size $2\times 2$ and is marked by a dashed blue line. The black dotes stand for the lattice sites which hosts the physical fields whereas the solid blue lines for the four virtual fields of the ingoing and outgoing direction to the block.}
		\label{Fig2}
	\end{figure}
	
	In fact, (\ref{2.1}) has dependence on $\nabla\phi$ too. We close this section by showing how this and generalization thereof is achieved. The local physical Hilbert space $\mathcal{H}_{\text{phys}}$ may be defined on a block of lattice sites rather on a single point, as illustrated in Fig. \ref{Fig2}.
	\begin{equation}
		\mathcal{H}_{phys}\left(\textbf{x}\right)=\underset{\hat{\mathbf{a}}\in\text{block}}{\bigotimes}\mathcal{H}_{\phi\left(\textbf{x}+\hat{\mathbf{a}}\right)}
	\end{equation}
	
	The larger the block, the higher derivative terms for the physical field $\phi$ that can be written. We now make the observation that the virtual and physical fields do not have to lay on the same lattice - the virtual fields after all connect between the physical Hilbert spaces which may contain more than one lattice site.

	It is interesting to note that this simple operation cannot be applied to the virtual fields since they are restricted to lay on the boundary of the block rather then inside its bulk. This fact seems to hinder the ability to construct higher derivatives for the virtual fields.

	\section{Symmetries}
	PEPS are well-known for the ease of describing symmetries with them: one can parameterize families of PEPS which will be invariant under some symmetry group \cite{sanz_matrix_2009,molnar_normal_2018,cirac2021matrix}.

	Consider, at first, the case of a complex scalar field. We can define a global $U(1)$ transformation on it, such that
	\begin{equation}
		U\left(\theta\right)\left|\left\{\phi\left(\mathbf{x}\right)\right\}\right\rangle = \left|\left\{e^{i\theta}\phi\left(\mathbf{x}\right)\right\}\right\rangle
	\end{equation}
	For any	$\theta \in \left[0,2\pi\right)$. If we wish our PEPS to be invariant under the symmetry operation, we will need to define some transformation rules for the virtual fields too, e.g.
	\begin{equation}
		v_{\alpha}\left(\mathbf{x}\right) \rightarrow e^{i\theta}v_{\alpha}\left(\mathbf{x}\right)
	\end{equation}
Changing the integration variables (the integration measure $\mathcal{D}\phi\left(\mathbf{x}\right)$ is invariant under a unitary change of variables), we will get that $U\left(\theta\right)\left|\psi\right\rangle = \left|\psi\right\rangle$ if 
\begin{equation}
	V\left[v_{\alpha},\nabla v_{\alpha},\phi\right]=V\left[e^{i\theta}v_{\alpha},e^{i\theta}\nabla v_{\alpha},e^{i\theta}\phi\right]
\end{equation}

If, for example, we construct a Gaussian cPEPS, this will be satisfied if we pick
	\begin{equation}
	\begin{aligned}
		\mathcal{A}=&-\frac{1}{4}Z_{\alpha\beta}\nabla \overline{v}_{\alpha}\left(\textbf{x}\right)\nabla v_{\beta}\left(\textbf{x}\right)-\frac{1}{4}A_{\alpha\beta}\overline{v}_{\alpha}\left(\textbf{x}\right)v_{\beta}\left(\textbf{x}\right) \\
		&+z_{\alpha}\nabla \overline{v}_{\alpha}\left(\textbf{x}\right)\nabla\phi\left(\textbf{x}\right)+a_{\alpha}\overline{v}_{\alpha}\left(\textbf{x}\right)\phi\left(\textbf{x}\right)-\frac{c}{4}\left|\phi\left(\textbf{x}\right)\right|^{2} + c.c
	\end{aligned}
\end{equation}
	
	This can also be achieved through the PEPS construction, by properly parametrizing the states $\left|A\left(\mathbf{x}\right)\right\rangle$ and $\left|L_i\left(\mathbf{x}\right)\right\rangle$ (see, e.g. \cite{zohar_building_2016}); but can also be done directly in the continuum.
	
	We can think of more general settings. Suppose that instead of the field $\phi\left(\mathbf{x}\right)$ we introduce a vector, or a multiplet of fields, either real or complex, $\left\{\phi_a\left(\mathbf{x}\right)\right\}_{a=1}^r$, transforming as some  representation $j$ of dimension $r$ of some group $G$; that is, for each group element $g \in G$, we define the unitary transformation
	\begin{equation}
	U\left(g\right)\left|\left\{\phi_a\left(\mathbf{x}\right)\right\}\right\rangle = \left|\left\{D^j_{ab}\left(g\right)\phi_b\left(\mathbf{x}\right)\right\}\right\rangle
	\end{equation}
where $D^j_{ab}\left(g\right)$ is the $j$ representation matrix of $g$ (of size $r\times r$).

If we wish to construct a cPEPS $\left|\psi\right\rangle$ with the symmetry property
\begin{equation}
U\left(g\right) \left|\psi\right\rangle = \left|\psi\right\rangle \quad\quad \forall g\in G
\end{equation}
we will need to use virtual fields that are also charged under this group, that is, the $v$ fields will have to carry an $a$ index, forming multiplets of the group $G$. One can use several multiplets and copies thereof, as long as the multiplets are fully included. The virtual and physical representations do not even have to be the same, as long as they are coupled properly. Thus, in general, the virtual fields transform as
\begin{equation}
	v_{\alpha,a}\left(\mathbf{x}\right) \rightarrow D^J_{ab}\left(g\right)v_{\alpha,b}\left(\mathbf{x}\right) 
\end{equation}
with some (irreducible or reducible) representation $J$.
Our cPEPS will be defined with
\begin{equation}
\mathcal{A}=-\frac{1}{2}Z_{\alpha\beta}\nabla_{i}\overline{v}_{\alpha,a}\nabla_{i} v_{\beta,a}+V\left[v_{\alpha,a},\nabla v_{\alpha,a},\phi_a\right]
\end{equation}
and the symmetry condition will be
\begin{equation}
	\begin{aligned}
	&V\left[v_{\alpha,a},\nabla v_{\alpha,a},\phi_a\right]=\\&V\left[D^J_{ab}\left(g\right)v_{\alpha,b},D^J_{ab}\left(g\right)\nabla  v_{\alpha,b},D^j_{ab}\left(g\right)\phi_b\right] 
	\end{aligned}
\end{equation}

which can also be simplified, in a straightforward manner, in the Gaussian case (it can also be seen as a continuum limit of the case studied in \cite{zohar_combining_2018}, when changing to scalar fields).

Finally, if we keep $\mathcal{A}$ real, we will also have a charge-conjugation symmetry in the case of complex fields.
	
	\section{Discussion}

	We have developed a continuous Projected Entangled Pair State (cPEPS) for quantum fields and explicitly shown how such a state can approach the free field theory vacuum as the bond dimension is increased. In addition we have tackled the question on whether this approximation has a regularization independent (universal) significance. To that end we have used the quantum fidelity as a measure of distinguishability between the cPEPS and the real physical state that it approximates, and found that it encapsulates a universal term that is independent on the short-scale behavior of the problem.
	
	Subsequently we have built the cPEPS in a bottom-up approach. Starting from the lattice, a PEPS paired with a unique ansatz was used to produce the wanted result in the continuum. A central feature of the cPEPS is that it enjoys a global symmetry under any desired symmetry group of interest, as well as translational and continuous rotational symmetries.
	
While we have focused on scalars, fields with higher spin may also be taken in consideration. A similar construction as the one done in section III is expected to be possible for physical and virtual fields with higher spin, albeit left beyond the scope of this paper. In fact, only two generalizations are required for spin 1 vector fields and spin $\frac{1}{2}$ fermions. Once they are laid down, higher integer or respectively half-integer spin fields can be constructed by taking tensor products and symmetrizing/anti-symmetrizing.
	
	Furthermore, an important benefit of this bottom-up construction is that tools and ideas that apply for PEPSs are transferred  to cPEPSs - section IV is an important example for that. The PEPS framework has been elevated to include physical gauge fields as well \cite{zohar_building_2016,zohar_combining_2018}, with a very clear prescription on how to do so. We expect that following through this procedure to the continuum will result in "minimal-coupling" of (\ref{3.8}) (in the same sense that an action is minimally coupled). Nevertheless we leave this to a more thorough investigation in the future.
	
	The fidelity was used for estimating the accuracy of the cPEPS approximation to the real, physical ground-state. In most cases however, the exact form of the ground-state is unknown, and hence it is the tensor-network state that serves as an ansatz that is designed to capture most of the ground-states important properties. If so, one immediate implementation of the cPEPS is to serve as a variational ansatz which is numerically adjusted to give a minimal expectation value for some  Hamiltonian of interest. To deal with the usual divergences that appear in the ground-state energy of quantum field theories, it was demonstrated to be efficient to minimize the normal-ordered/renormalized Hamiltonian instead of the bare one \cite{tilloy2021relativistic},\cite{PhysRevD.91.085011}. If a different renormalization scheme would be used, will a smaller bond dimension be able to yield a similar result? In our non-interacting example, we have found that indeed there is a universal term (which does not depend on the cutoff) appearing in the fidelity per-site, yet the larger picture remains open for future study.
	
	Along with this, there is another context to which the cPEPS can be closely related. A tensor-network state can be thought of as a representative of a class of states with a specific structure of symmetry and entanglement. Since the critical properties of a phase transitions is normally dependent upon the robust features of the two phases, one can build tensor-network states which lay in the universality class of each phase and in that way "engineer" a quantum phase transition \cite{PhysRevLett.97.110403}. There is then great merit in calculating the fidelity of two nonidentical cPEPSs belonging to different symmetry classes - which can be easily created with the tools portrayed in this paper.
	
	Fidelity of many-body quantum states as a probe for quantum phase transitions has been a popular source of research (see \cite{doi:10.1142/S0217979210056335} and references therein). The quantum fidelity has proven to be a useful analytical tool for gaining insight about the characteristics of the phase transition and it has a unique critical behavior around the transition point on its own. Exact form of the fidelity is in general not easily calculable, and known examples exist in large for integrable models in where the ground-state has a known form. We therefore hope that the cPEPS together with standard perturbative techniques can open a way to a new class of phase transitions that can be studied via the fidelity approach and we leave that to future work.
	
	\begin{acknowledgements}
	The authors would like to thank Michael Smolkin and Patrick Emonts for helpful and inspiring discussions. This research was supported by the Israel Science Foundation (grant No. 523/20).
	\end{acknowledgements}
	
	\appendix

	\section{Equivalence with CTNS}\label{appendixctns}
In \cite{PhysRevX.9.021040}, the following continuous tensor network states (CTNS)  were defined for real scalar fields (rewritten for infinite spaces and using our notation conventions):
\begin{widetext}\label{CTNS}
	\begin{equation}
		\left|CTNS\right\rangle=\int \mathcal{D} v \exp \left(-\int d^d x \left[\frac{1}{2}\underset{\alpha=1}{\overset{D}{\sum}}\left(\nabla v_{\alpha}\left(\mathbf{x}\right)\right)^2
		+V\left[\left\{v_{\alpha}\right\}\right]-f\left[\left\{v_{\alpha}\right\}\right]\Psi^{\dagger}\left(\mathbf{x}\right)\right]
		\right)\left|0\right\rangle
	\end{equation}
	where $V,f$ are some functionals of the virtual fields, and $\Psi^{\dagger}\left(\mathbf{x}\right)$ is a local field (bosonic) creation operator, defined at $\mathbf{x}$, as
	\begin{equation}
		\Psi^{\dagger}\left(\mathbf{x}\right) = \frac{1}{\sqrt{2}}\left(\Phi\left(\mathbf{x}\right)-i\Pi\left(\mathbf{x}\right)\right)
	\end{equation}
	and $\left|0\right\rangle$ is the Fock vacuum annihilated by all $\Psi\left(\mathbf{x}\right)$, rather than the ground state of some relativistic (even free) field theory, such as what we denote in the main text by $\left|\Omega_f\right\rangle$. For this reason, we can also write
	\begin{equation}
		\left|CTNS\right\rangle=\int \mathcal{D} v \exp \left(-\int d^d x \left[\frac{1}{2}\underset{\alpha=1}{\overset{D}{\sum}}\left(\nabla v_{\alpha}\left(\mathbf{x}\right)\right)^2
		+V\left[\left\{v_{\alpha}\right\}\right]-f\left[\left\{v_{\alpha}\right\}\right]\Psi^{\dagger}\left(\mathbf{x}\right)
		\right]\right)
		\exp\left(\int d^d xf\left[\left\{v_{\alpha}\right\}\right]\Psi\left(\mathbf{x}\right)\right)
		\left|0\right\rangle
	\end{equation}
	Furthermore, thanks to the canonical commutation relation of Eq. (\ref{canrel}), we find that
	\begin{equation}
		\left[\Psi\left(\mathbf{x}\right),\Psi^{\dagger}\left(\mathbf{y}\right)\right]=\delta^{(d)}\left(\mathbf{x}-\mathbf{y}\right)
	\end{equation}
	and thus
	\begin{equation}
		\left|CTNS\right\rangle=
		\int \mathcal{D} v \exp \left(-\int d^d x \left[\frac{1}{2}\underset{\alpha=1}{\overset{D}{\sum}}\left(\nabla v_{\alpha}\left(\mathbf{x}\right)\right)^2
		+V\left[\left\{v_{\alpha}\right\}\right]+\frac{1}{2}f^2\left[\left\{v_{\alpha}\right\}\right]-f\left[\left\{v_{\alpha}\right\}\right]\left(\Psi^{\dagger}\left(\mathbf{x}\right)+\Psi\left(\mathbf{x}\right)\right)\right]
		\right)
		\left|0\right\rangle
	\end{equation} 
	Next, using 
	\begin{equation}
		\Phi\left(\mathbf{x}\right) = \frac{1}{\sqrt{2}}\left(\Psi^{\dagger}\left(\mathbf{x}\right)+\Psi\left(\mathbf{x}\right)\right),
	\end{equation}
	\begin{equation}
		\mathbb{I}=\int \mathcal{D} \phi \left|\left\{\phi\left(\mathbf{x}\right)\right\}\right\rangle
		\left\langle\left\{\phi\left(\mathbf{x}\right)\right\}\right|
	\end{equation}
	(with field operators eigenstates) and 
	\begin{equation}
		\left\langle\left\{\phi\left(\mathbf{x}\right)\right\}|0\right\rangle \propto 
		\exp \left(-\frac{1}{2}\int d^d x \phi^2\left(\mathbf{x}\right) \right)
	\end{equation}
	we finally obtain
	\begin{equation}
		\left|CTNS\right\rangle=
		\int \mathcal{D}\phi \mathcal{D} v \exp \left(-\int d^d x \left[\frac{1}{2}\underset{\alpha=1}{\overset{D}{\sum}}\left(\nabla v_{\alpha}\left(\mathbf{x}\right)\right)^2
		+V\left[\left\{v_{\alpha}\right\}\right]+\frac{1}{2}f^2\left[\left\{v_{\alpha}\right\}\right]-\sqrt{2}f\left[\left\{v_{\alpha}\right\}\right]\phi\left(\mathbf{x}\right)+\frac{1}{2}\phi^2\left(\mathbf{x}\right)\right]
		\right)
		\left|\left\{\phi\left(\mathbf{x}\right)\right\}\right\rangle
	\end{equation} 
- that is, the CTNS expressed in a cPEPS form.

The other direction is true as well:  it is also possible to express any cPEPS (\ref{2.1}) in the CTNS form (\ref{CTNS}). To do that, we recall the definition of field coherent states,
	\begin{equation}
	\left|\left\{\alpha\left(\mathbf{x}\right)\right\}\right\rangle =e^{\int d^{d}x\,\alpha\left(\mathbf{x}\right)\Psi^{\dagger}\left(\mathbf{x}\right)}\left|0\right\rangle
\end{equation}
-  eigenstates of the field annihilation operator $\Psi\left(\mathbf{x}\right)$, with the overcompleteness relation
\begin{equation}\label{overc}
\mathbb{I}=\int\mathcal{D}\alpha\left(\mathbf{x}\right)
\mathcal{D}\overline{\alpha}\left(\mathbf{x}\right) e^{-\int d^{d}\mathbf{x}\left|\alpha\left(\mathbf{x}\right)\right|^{2}}\left|\left\{\alpha\left(\mathbf{x}\right)\right\}\right\rangle \left\langle \left\{\alpha\left(\mathbf{x}\right)\right\}\right|.
\end{equation}

The overlap between a coherent state and a field configuration state is given by
\begin{equation}
	\begin{aligned}
	\left\langle \left\{\alpha\left(\mathbf{x}\right)\right\}\middle| \left\{\phi\left(\mathbf{x}\right)\right\}\right\rangle& =
	\left\langle 0 \right|  e^{\int d^dx \overline{\alpha}\left(\mathbf{x}\right) \Psi\left(\mathbf{x}\right)} \left|\left\{\phi\left(\mathbf{x}\right)\right\}\right\rangle
	=	\left\langle 0 \right| 
	 e^{\int d^dx \overline{\alpha}\left(\mathbf{x}\right) \Psi^{\dagger}\left(\mathbf{x}\right)}
	 e^{\int d^dx \overline{\alpha}\left(\mathbf{x}\right) \Psi\left(\mathbf{x}\right)} \left|\left\{\phi\left(\mathbf{x}\right)\right\}\right\rangle
	 \\&=
	 e^{-\int d^dx\left(\frac{1}{2}\overline{\alpha}^2\left(\mathbf{x}\right)-\sqrt{2}\overline{\alpha}\left(\mathbf{x}\right)\phi\left(\mathbf{x}\right)\right)}\left\langle 0 \middle| \left\{\phi\left(\mathbf{x}\right)\right\}\right\rangle \propto 
	 e^{-\int d^dx\left(\frac{1}{2}\phi^2\left(\mathbf{x}\right)+\frac{1}{2}\overline{\alpha}^2\left(\mathbf{x}\right)-\sqrt{2}\overline{\alpha}\left(\mathbf{x}\right)\phi\left(\mathbf{x}\right)\right)}
	 \end{aligned}
\end{equation}
Along with the overcompleteness relation (\ref{overc}), this gives rise to
\begin{equation}
	\begin{aligned}
\left|\left\{\phi\left(\mathbf{x}\right)\right\}\right\rangle
&\propto \int\mathcal{D}\alpha\left(\mathbf{x}\right)
\mathcal{D}\overline{\alpha}\left(\mathbf{x}\right) 
 e^{-\int d^dx\left(\frac{1}{2}\phi^2\left(\mathbf{x}\right)+\frac{1}{2}\overline{\alpha}^2\left(\mathbf{x}\right)-\sqrt{2}\overline{\alpha}\left(\mathbf{x}\right)\phi\left(\mathbf{x}\right)
 	+\overline{\alpha}\left(\mathbf{x}\right)\alpha\left(\mathbf{x}\right)\right)} 	
 \left|\left\{\alpha\left(\mathbf{x}\right)\right\}\right\rangle
 \\&=
 \int\mathcal{D}\alpha\left(\mathbf{x}\right)
 \mathcal{D}\overline{\alpha}\left(\mathbf{x}\right) 
 e^{-\int d^dx\left(\frac{1}{2}\phi^2\left(\mathbf{x}\right)+\frac{1}{2}\overline{\alpha}^2\left(\mathbf{x}\right)-\sqrt{2}\overline{\alpha}\left(\mathbf{x}\right)\phi\left(\mathbf{x}\right)
 	+\overline{\alpha}\left(\mathbf{x}\right)\alpha\left(\mathbf{x}\right)\right)} 	
 e^{\int d^{d}x\,\alpha\left(\mathbf{x}\right)\Psi^{\dagger}\left(\mathbf{x}\right)}\left|0\right\rangle
  \end{aligned}
\end{equation}
Plugging this into Eq. (\ref{2.1}), we obtain a CTNS form for the CPEPS, introducing three more virtual fields, $\alpha\left(\mathbf{x}\right)$, $\overline{\alpha}\left(\mathbf{x}\right)$, $\phi\left(\mathbf{x}\right)$. This completes the proof of equivalence between the states.
\end{widetext}

	\section{Continued Fraction Approximation}
	\label{appcfrac}
	
	Hereby we shall introduce a specific and simpler choice of the more general Gaussian cPEPS (\ref{2.1}) which enables the presentation of $\omega_{D}\left(k\right)$ in the form of a continued fraction. It allows for a rather elegant representation of the ground-state of a free field theory. Furthermore, continued fractions in general have many appealing properties, allowing one to gain more knowledge about the convergence. Given that, we  narrow down to the parameter choice of Eq. (\ref{params}):
\begin{equation}
			A_{\alpha\beta}+Z_{\alpha\beta}k^{2}=2m\delta_{\alpha,\beta} \begin{cases} 1	 & \alpha \space \text{ even} \\
				k^2 & \alpha \space \text{ odd} \end{cases} + \frac{k^2}{\sqrt{2}}\left(\delta_{\alpha,\beta-1}+\delta_{\alpha,\beta+1}\right)
	\end{equation}	
	\begin{equation}\label{A.1}
			=
			\begin{pmatrix}2mk^{2} & k^{2}/\sqrt{2}\\
				k^{2}/\sqrt{2} &2 m & k^{2}/\sqrt{2}\\
				& k^{2}/\sqrt{2} & 2mk^{2} & k^{2}/\sqrt{2}\\
				&  & k^{2}/\sqrt{2} & \ddots & k^{2}/\sqrt{2}\\
				&  &  & k^{2}/\sqrt{2} & 2m
			\end{pmatrix}
			\end{equation}
and $a_{\alpha}=0,z_{\alpha}=\sqrt{2}\delta_{\alpha,1},c=m$.
	
	We first need to take care of the fact that there is more than one way of fixing the mass dimension $\left[\chi\right]$ of the virtual fields $\chi_{\alpha}$. We find it useful to choose the dimension to be the same as that of the physical field $\phi$: $\left[\chi\right]=\left[\phi\right]=\frac{d-1}{2}$. Recall that the dimension of the physical field $\left[\phi\right]$ is completely set by the Klein-Gordon action. Under this convention the dimensions of the parameters defined in (\ref{2.1}) are as follows:
	\begin{equation}
		[c]=1\;,\;[A_{\alpha\beta}]=[a_{\alpha}]=1\;,\;[Z_{\alpha\beta}]=[z_{\alpha}]=-1.
	\end{equation}
	Notice that we have set the physical mass $m$ which appears in $\omega_f$ as a single energy scale and used it to fix the dimensions of all of the rest of the parameters. Thus $B_{\alpha},Z_{\alpha}$
	and $A_{\alpha}$ are dimensionless. Integrating out the virtual fields (as we show below) results in:
		\begin{equation}\label{A.2}
		\begin{array}{cc}
			\omega_{D}\left(k\right)=m+\cfrac{k^{2}}{2m+\cfrac{k^{2}}{2m+\cfrac{k^{2}}{2m+...\cfrac{k^{2}}{2m+\cfrac{k^{2}}{2m}}}}}
			\\
			\phantom{A}
		\end{array}
	\end{equation}
	In this form it is a rather simple task to make the cPEPS (\ref{A.1}) approach the vacuum state of a free field theory $\left|\Omega_{f}\right\rangle$:
	
	\begin{equation}
		\left|\Omega_{f}\right\rangle=\int\mathcal{D}\phi\left(\textbf{k}\right)e^{-\frac{1}{2}\omega_{f}\left(k\right)\bar{\phi}\left(\textbf{k}\right)\phi\left(\textbf{k}\right)}\left|\left\{\phi\left(\textbf{k}\right)\right\} \right\rangle
	\end{equation}
	
	$\omega_{f}\left(k\right)=\sqrt{k^2+m^2}$ has a continued fraction representation:
		\begin{equation}
		\omega_{f}\left(k\right)=m+\cfrac{k^{2}}{2m+\cfrac{k^{2}}{2m+\frac{k^{2}}{\ddots}}}
	\end{equation}
$\omega_{D}\left(k\right)$  approaches
	$\omega_{f}\left(k\right)$ as $D$ grows. The convergence however is not uniform as the remainder diverges as $k$ increases. This can be seen by the fact that $\omega_{D}\left(k\right)$ asymptotically behaves for large values of $k$ as $k^{2}$ or as a constant depending if the truncation of the continued fraction is even or odd, whereas $\omega_{f}\left(k\right)$ asymptotically behaves as $k$.
	
	In fact, the continued fraction representation is not unique. This is  not surprising given the fact that so is the more general Padé approximant. 
	
Eq. (\ref{A.2}) is obtained inductively. To that end we integrate out the virtual fields one by one, and it is useful to present the following notation:
		\begin{equation}\label{A.3}
		\begin{aligned}
		&\left|\psi\right\rangle 	=\int\mathcal{D}\phi\left(\textbf{k}\right)\overset{D}{\underset{\alpha =1}{\prod}}\mathcal{D}v_{\alpha}\left(\textbf{k}\right)
			e^{\int\frac{d^{d}k}{\left(2\pi\right)^{d}}\mathcal{A}_{D}\left[v_{\alpha}\left(\textbf{k}\right),\phi\left(\textbf{k}\right)\right]}\left|\left\{ \phi\left(\textbf{k}\right)\right\} \right\rangle
			\\
			& =\int\mathcal{D}\phi\left(\textbf{k}\right)\overset{D-1}{\underset{\alpha =1}{\prod}}\mathcal{D}v_{\alpha}\left(\textbf{k}\right)
			e^{\int\frac{d^{d}k}{\left(2\pi\right)^{d}}\mathcal{A}_{D-1}\left[v_{\alpha}\left(\textbf{k}\right),\phi\left(\textbf{k}\right)\right]}\left|\left\{ \phi\left(\textbf{k}\right)\right\} \right\rangle
			\\
			&=...
		\end{aligned}
	\end{equation}
	
	For obtaining (\ref{A.2}) we start by integrating $v_{\alpha =D}\left(k\right)$. To clarify that, we first write (\ref{A.3}) in a form where $v_{\alpha =D}\left(k\right)$ is separated from the rest of the virtual fields:
	\begin{equation}
	\begin{aligned}
		\mathcal{A}_{D}=&-\frac{2m}{2}\left|v_{D}\left(k\right)\right|^{2}\\
		&+
		\frac{k^{2}}{\sqrt{2}}
		\left(\bar{v}_{D}\left(k\right)v_{D-1}\left(k\right)+\bar{v}_{D-2}\left(k\right)v_{D}\left(k\right)\right) \\
		&		-\frac{1}{2}\underset{\alpha,\beta}{\overset{D-1}{\sum}}\left(Z+Ak^{2}\right)_{\alpha\beta}\bar{v}_{\alpha}\left(k\right)v_{\beta}\left(k\right) \\
		&+\frac{k^{2}}{\sqrt{2}}
		\left(\bar{v}_{1}\left(k\right)\phi\left(k\right)+\bar{\phi}\left(k\right)v_{1}\left(k\right)\right) -\frac{m}{2}\left|\phi\left(\textbf{k}\right)\right|^{2}.
	\end{aligned}
	\end{equation}
	Integrating out $v_{D}$ leads to:
	\begin{equation}
	\begin{aligned}
		\mathcal{A}_{D-1}=&-\frac{1}{2}\left(2mk^2+\frac{k^4}{2m}\right)\left|v_{D-1}\left(k\right)\right|^{2} \\
		&+
		\frac{k^{2}}{\sqrt{2}}
		\left(\bar{v}_{D-1}\left(k\right)v_{D-2}\left(k\right)+\bar{v}_{D-1}\left(k\right)v_{D-2}\left(k\right)\right) \\
	&		-\frac{1}{2}\underset{\alpha,\beta}{\overset{D-2}{\sum}}\left(Z+Ak^{2}\right)_{\alpha\beta}\bar{v}_{\alpha}\left(k\right)v_{\beta}\left(k\right) \\
	&+\frac{k^{2}}{\sqrt{2}}
	\left(\bar{v}_{1}\left(k\right)\phi\left(k\right)+\bar{\phi}\left(k\right)v_{1}\left(k\right)\right) -\frac{m}{2}\left|\phi\left(\textbf{k}\right)\right|^{2}.
	\end{aligned}
	\end{equation}
		Next, integrating  $v_{D-1}$ out leads to:
	\begin{equation}
	\begin{aligned}
		\mathcal{A}_{D-2}=&-\frac{1}{2}\left(2m+\frac{k^2}{2m + \frac{k^2}{2m}}\right)\left|v_{D-1}\left(k\right)\right|^{2} \\
		&+
		\frac{k^{2}}{\sqrt{2}}
		\left(\bar{v}_{D-1}\left(k\right)v_{D-2}\left(k\right)+\bar{v}_{D-1}\left(k\right)v_{D-2}\left(k\right)\right) \\
		&		-\frac{1}{2}\underset{\alpha,\beta}{\overset{D-2}{\sum}}\left(Z+Ak^{2}\right)_{\alpha\beta}\bar{v}_{\alpha}\left(k\right)v_{\beta}\left(k\right) \\
		&+\frac{k^{2}}{\sqrt{2}}
		\left(\bar{v}_{1}\left(k\right)\phi\left(k\right)+\bar{\phi}\left(k\right)v_{1}\left(k\right)\right) -\frac{m}{2}\left|\phi\left(\textbf{k}\right)\right|^{2}.
		\end{aligned}
	\end{equation}
The emergence of the continued fraction after integrating out all the (even) $D$ virtual fields can be clearly seen.

Note that the above derivation holds for $k \neq 0$; in the $k=0$ case, we see that all the virtual fields are decoupled from one another and from the physical field, and separate integrations will give rise to the expected result.
	
	\section{A Note on Parent Hamiltonians}
	
	A natural question put in the context of tensor-networks is about the properties and uniqueness (or non-uniqueness) of a local parent Hamiltonian $H_{D}$, whose ground state is the tensor network state - in our case, (\ref{2.2}). Since (\ref{2.2}) is a Gaussian state there must exist a quadratic parent Hamiltonian, and in fact there is a family of these with the following form:
	\begin{equation}\label{2.9}
		\begin{aligned}
			H_{D}\left(\phi,\pi\right)=\frac{1}{2}\int \frac{d^{d}k}{\left(2\pi\right)^{d}}\Big( &a\left(\textbf{k}\right)\bar{\Pi}\left(\textbf{k}\right)\Pi\left(\textbf{k}\right) \\
			&+b\left(\textbf{k}\right)\bar{\phi}\left(\textbf{k}\right)\phi\left(\textbf{k}\right)\Big)
		\end{aligned}
	\end{equation}
	where $\omega^{2}_{D}\left(\textbf{k}\right)=\frac{b\left(\textbf{k}\right)}{a\left(\textbf{k}\right)}$.
	In this aspect, $\omega_{D}\left(k\right)$ may be interpreted
	as a dispersion relation, obviously depnding on the bond dimension $D$ and the parameters which appear in the Gaussian cPEPS (\ref{2.3}). 
	
	All the Hamiltonians (\ref{2.9}) have the same spectrum as they are canonically equivalent to each other by such transformations which preserve the locality. However, since we would like the parent Hamiltonian \ref{2.9} to be local, we narrow down to cases where the functions $a\left(\textbf{k}\right),b\left(\textbf{k}\right)$ are finite polynomials of $\textbf{k}$. 
	
	\bibliography{ref}
\end{document}